\PassOptionsToPackage{unicode}{hyperref}
\PassOptionsToPackage{hyphens}{url}
\PassOptionsToPackage{dvipsnames,svgnames,x11names}{xcolor}
\documentclass[
]{article}
\usepackage{amsmath,amssymb}
\usepackage{lmodern}
\usepackage{iftex}
\ifPDFTeX
  \usepackage[T1]{fontenc}
  \usepackage[utf8]{inputenc}
  \usepackage{textcomp} 
\else 
  \usepackage{unicode-math}
  \defaultfontfeatures{Scale=MatchLowercase}
  \defaultfontfeatures[\rmfamily]{Ligatures=TeX,Scale=1}
\fi
\IfFileExists{upquote.sty}{\usepackage{upquote}}{}
\IfFileExists{microtype.sty}{
  \usepackage[]{microtype}
  \UseMicrotypeSet[protrusion]{basicmath} 
}{}
\makeatletter
\@ifundefined{KOMAClassName}{
  \IfFileExists{parskip.sty}{%
    \usepackage{parskip}
  }{
    \setlength{\parindent}{0pt}
    \setlength{\parskip}{6pt plus 2pt minus 1pt}}
}{
  \KOMAoptions{parskip=half}}
\makeatother
\usepackage{xcolor}
\usepackage{graphicx}
\makeatletter
\def\maxwidth{\ifdim\Gin@nat@width>\linewidth\linewidth\else\Gin@nat@width\fi}
\def\maxheight{\ifdim\Gin@nat@height>\textheight\textheight\else\Gin@nat@height\fi}
\makeatother
\setkeys{Gin}{width=\maxwidth,height=\maxheight,keepaspectratio}
\makeatletter
\def\fps@figure{htbp}
\makeatother
\NewDocumentCommand\citeproctext{}{}
\NewDocumentCommand\citeproc{mm}{%
  \begingroup\def\citeproctext{#2}\cite{#1}\endgroup}
\makeatletter
 \let\@cite@ofmt\@firstofone
 \def\@biblabel#1{}
 \def\@cite#1#2{{#1\if@tempswa , #2\fi}}
\makeatother
\newlength{\cslhangindent}
\newlength{\csllabelwidth}
\newenvironment{CSLReferences}[2] 
 {\begin{list}{}{%
  \setlength{\itemindent}{0pt}
  \setlength{\leftmargin}{0pt}
  \setlength{\parsep}{0pt}
  \ifodd #1
   \setlength{\leftmargin}{\cslhangindent}
   \setlength{\itemindent}{-1\cslhangindent}
  \fi
  \setlength{\itemsep}{#2\baselineskip}}}
 {\end{list}}
\usepackage{calc}

\ifLuaTeX
\usepackage[bidi=basic]{babel}
\else
\usepackage[bidi=default]{babel}
\fi
\babelprovide[main,import]{american}

\def\languageshorthands#1{}
\ifLuaTeX
  \usepackage{selnolig}  
\fi
\IfFileExists{bookmark.sty}{\usepackage{bookmark}}{\usepackage{hyperref}}
\IfFileExists{xurl.sty}{\usepackage{xurl}}{} 
\hypersetup{
  pdftitle={popclass: a python package for classifying microlensing
events},
  pdfauthor={Greg Sallaberry, Zofia Kaczmarek, Peter McGill, Scott E.
Perkins, William A. Dawson, Caitlin G. Begbie},
  pdflang={en-US},
  colorlinks=true,
  linkcolor={Maroon},
  filecolor={Maroon},
  citecolor={Blue},
  urlcolor={Blue},
  pdfcreator={LaTeX via pandoc}}

\title{popclass: a python package for classifying microlensing events}

\definecolor{c53baa1}{RGB}{83,186,161}
\definecolor{c202826}{RGB}{32,40,38}

\usepackage[affil-it]{authblk}
\usepackage{orcidlink}
\author[1%
  *%
  ]{Greg Sallaberry%
    \,\orcidlink{0009-0001-4859-5205}\,%
    }
\author[1,2%
  *%
  ]{Zofia Kaczmarek%
    \,\orcidlink{0009-0007-4089-5012}\,%
    }
\author[1%
  *%
  \ensuremath\mathparagraph]{Peter McGill%
    \,\orcidlink{0000-0002-1052-6749}\,%
    }
\author[1%
  *%
  ]{Scott E. Perkins%
    \,\orcidlink{0000-0002-5910-3114}\,%
    }
\author[1%
  ]{William A. Dawson%
    \,\orcidlink{0000-0003-0248-6123}\,%
    }
\author[3%
  ]{Caitlin G. Begbie%
    \,\orcidlink{0009-0006-8866-4224}\,%
    }

\affil[1]{Space Science Institute, Lawrence Livermore National
Laboratory, 7000 East Ave., Livermore, CA 94550, USA%
  }
\affil[2]{Zentrum für Astronomie der Universität Heidelberg,
Astronomisches Rechen-Institut, Mönchhofstr. 12-14, 69120 Heidelberg,
Germany%
  }
\affil[3]{University of California, Berkeley, Astronomy Department,
Berkeley, CA 94720, USA%
  }
\affil[$\mathparagraph$]{Corresponding author: %
}
\affil[*]{These authors contributed equally.}
\date{17 October 2024}

\begin{document}
\maketitle

\section{Summary}\label{summary}

\texttt{popclass} is a python package that provides a flexible,
probabilistic framework for classifying the lens of a gravitational
microlensing event. Gravitational microlensing occurs when a massive
foreground object (e.g., a star, white dwarf or a black hole) passes in
front of and lenses the light from a distant background star. This
causes an apparent brightening, and shift in position, of the background
source. In most cases, characteristics of the microlensing signal do not
contain enough information to definitively identify the lens type.
However, different lens types can lie in different but overlapping
regions of the characteristics of the microlensing signal. For example,
black holes tend to be more massive than stars and therefore cause
microlensing signals that are longer. Current Galactic simulations allow
the prediction of where different lens types lie in the observational
space and can therefore be leveraged to classify events (e.g.,
\citeproc{ref-Lam2020}{Lam et al., 2020}).

\texttt{popclass} allows a user to match characteristics of a
microlensing signal to a simulation of the Galaxy to calculate lens type
probabilities for an event (see Figure 1). Constraints on any
microlensing signal characteristics and any Galactic model can be used.
\texttt{popclass} comes with an interface to \texttt{ArviZ}
(\citeproc{ref-arviz_2019}{Kumar et al., 2019}) and \texttt{pymultinest}
(\citeproc{ref-Buchner2016}{Buchner, 2016}) for microlensing signal
constraints, pre-loaded Galactic models, plotting functionality, and
classification uncertainty quantification methods. The probabilistic
framework for popclass was developed in Perkins et al.
(\citeproc{ref-Perkins2024}{2024}), used in Fardeen et al.
(\citeproc{ref-Fardeen2024}{2024}) and has been applied to classifying
events in Kaczmarek et al. (\citeproc{ref-Kaczmarek2024}{2024}).

\section{Statement of need}\label{statement-of-need}

The advent of the Vera C. Rubin Observatory
(\citeproc{ref-Ivezic2019}{Ivezić et al., 2019}) and the \emph{Nancy
Grace Roman Space Telescope} (\citeproc{ref-Spergel2015}{Spergel et al.,
2015}) will provide tens-of-thousands of microlensing events per year
(e.g., \citeproc{ref-Abrams2023}{Abrams et al., 2023};
\citeproc{ref-Penny2019}{Penny et al., 2019}). To maximize the science
output of this event stream it is critical to identify events that have
a high probability of being caused by a lens type such as a black hole
(\citeproc{ref-Lam2022}{Lam et al., 2022}; \citeproc{ref-Sahu2022}{Sahu
et al., 2022}), and then to allocate expensive follow-up observations
such as space-based astrometry (e.g., \citeproc{ref-Sahu2022}{Sahu et
al., 2022}) or ground-based adaptive optics imaging (e.g.,
\citeproc{ref-Terry2022}{Terry et al., 2022}) to confirm their nature.

Current microlensing software packages such as \texttt{DarkLensCode}
(\citeproc{ref-Howil2024}{Howil et al., 2024}) or \texttt{PyLiMASS}
(\citeproc{ref-Bachelet2024}{Bachelet et al., 2024}) estimate lens
mass-distance constraints using microlensing event light curve and
additional auxiliary information (e.g., source proper motions,
distances, color, or finite source effects). Using auxiliary information
makes these current methods powerful but limits them to only be
effective for events with the available auxiliary data. Moreover, no
current software tools explicitly predict lens type, and they always
assume a fixed Galactic model. \texttt{popclass} fills the need for a
flexible microlensing classification software package that can be
broadly applied to classify all events from the Vera C. Rubin
Observatory and \emph{Nancy Grace Roman Space Telescope} and can be used
with any Galactic model in the form of a simulation.

\section{Method}\label{method}

\texttt{popclass} relies on the Bayesian classification framework in
(\citeproc{ref-Perkins2024}{Perkins et al., 2024}). Consider the data
from a single microlensing event \(\boldsymbol{d}\), and a model of the
Galaxy \(\mathcal{G}\). \texttt{popclass} calculates the probability
that the lens of the event belongs to each lens class,
\(\text{class}_L\), where \(\text{class}_L\in\text{classes}\) and, for
example,
\(\text{classes} = \{\text{Star, Neutron Star, White Dwarf, Black Hole}\}\).
Namely, \texttt{popclass} calculates

\[p(\text{class}_L| \boldsymbol{d}, \mathcal{G}) \text{ for } \text{class}_L\in\text{classes}.\]

Using Bayes' theorem,

\[p(\text{class}_L| \boldsymbol{d}, \mathcal{G}) = \frac{p(\text{class}_L| \mathcal{G})p(\boldsymbol{d}| \text{class}_L, \mathcal{G})}{p(\boldsymbol{d}| \mathcal{G})}.\]

Assuming that the set of considered lens classes is complete,
\(p(\boldsymbol{d}| \mathcal{G})\) is a normalization factor such that
all lens class probabilities sum to unity. Using importance sampling
(e.g., \citeproc{ref-Hogg2010}{Hogg et al., 2010}) with \(S\)
independent posterior samples
\(\theta_{c}\sim p(\theta|\boldsymbol{d})\) drawn under some prior,
\(\pi(\theta)\), obtained from fitting some set of microlensing signal
parameters, \(\theta\),

\[p(\text{class}_L | \boldsymbol{d}, \mathcal{G}) = \frac{p(\text{class}_L| \mathcal{G})}{p(\boldsymbol{d}| \mathcal{G})}
    \times \frac{1}{S} \sum _{c=0}^{S} \frac{p(\theta _c | \text{class}_L, \mathcal{G})}{\pi(\theta _{c})}\].

This allows the use of previously calculated posterior samples to
perform lens classification for a single event in the context of a
Galactic model. The term \(p(\theta_c | \text{class}_ L, \mathcal{G})\)
is calculated using kernel density estimation in \texttt{popclass} over
\(\theta\) with a simulated catalog of microlensing events from
\(\mathcal{G}\). \(p(\text{class}_L | \mathcal{G})\) is the prior
probability that an event belongs to each class before any data is seen,
which is just set by relative number of expected events predicted by the
Galactic model \(\mathcal{G}\).

\begin{figure}
\centering
\includegraphics{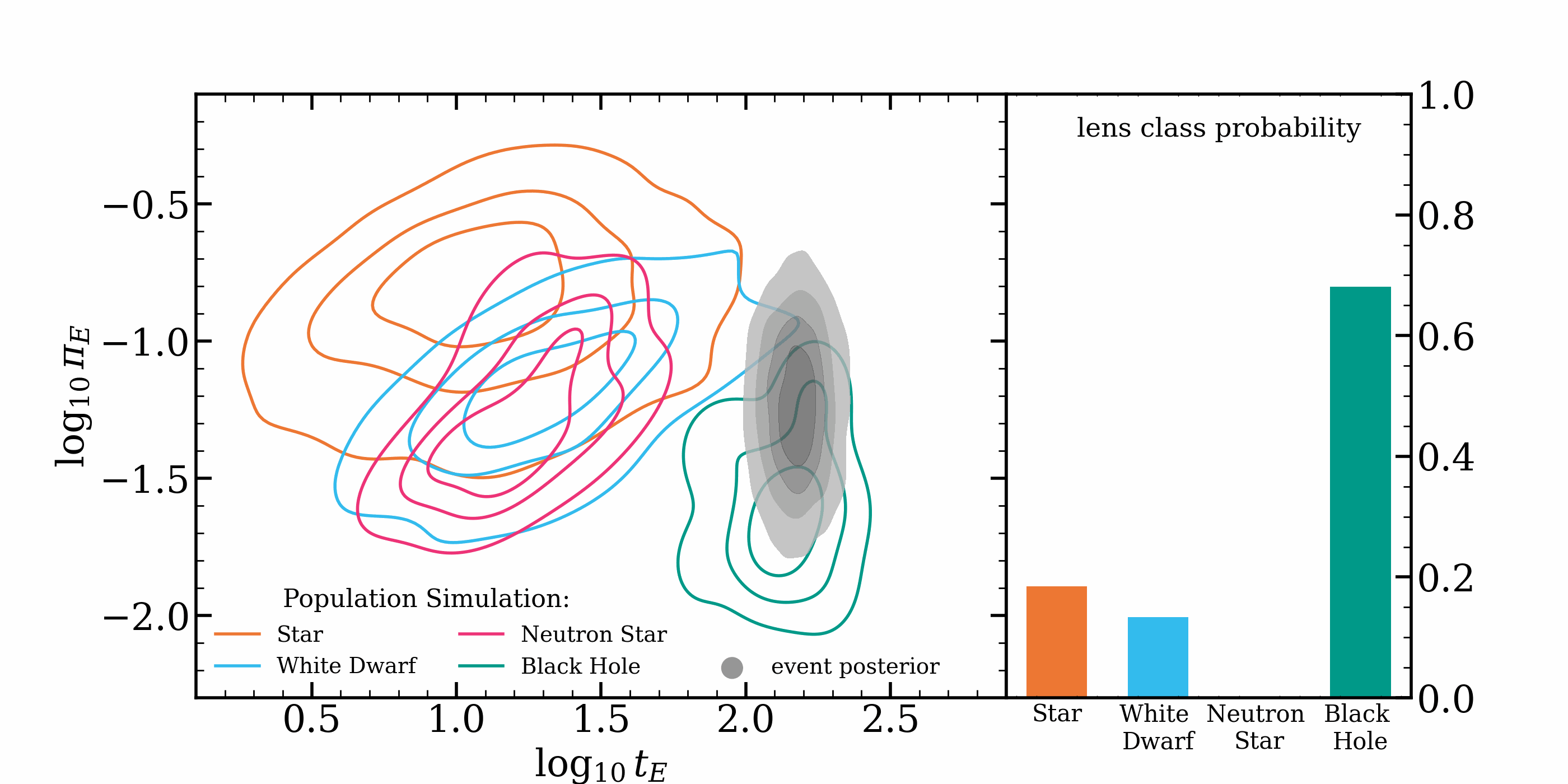}
\caption{Left: posterior distribution of an event in
log10(timescale)-log10(parallax) space, overlaid on `star', `white
dwarf', `neutron star' and `black hole' contours. Right: bars showing
probabilities of that event belonging to each of the lens populations.}
\end{figure}

\section{Acknowledgements}\label{acknowledgements}

\texttt{popclass} depends on numpy (\citeproc{ref-Harris2020}{Harris et
al., 2020}), scipy (\citeproc{ref-Virtanen2020}{Virtanen et al., 2020}),
asdf (\citeproc{ref-Greenfield2015}{Greenfield et al., 2015}),
matplotlib (\citeproc{ref-Hunter2007}{Hunter, 2007}), and scikit-learn
(\citeproc{ref-sklearn_api}{Buitinck et al., 2013}). This work was
performed under the auspices of the U.S. Department of Energy by
Lawrence Livermore National Laboratory (LLNL) under Contract
DE-AC52-07NA27344. The document number is LLNL-JRNL-870290 and the code
number is LLNL-CODE-2000456. The theoretical foundation of this work was
established under support from Lawrence Livermore National Laboratory's
Laboratory Directed Research and Development Program under project
22-ERD-037. The software implementation for this project was funded
under the LLNL Space Science Institute's Institutional Scientific
Capability Portfolio funds in partnership with LLNL's Academic
Engagement Office. ZK acknowledges support from the 2024 LLNL Data
Science Summer Institute and is a Fellow of the International Max Planck
Research School for Astronomy and Cosmic Physics at the University of
Heidelberg (IMPRS-HD).

\section*{References}\label{references}
\addcontentsline{toc}{section}{References}

\phantomsection\label{refs}

\end{document}